\documentclass[twocolumn,showpacs,preprintnumbers,amsmath,amssymb]{revtex4} 

\usepackage{graphicx}
\usepackage{dcolumn}
\usepackage{bm}

\renewcommand{\d}{{\rm d}} 

\begin{document} 
 
 
\title{Coulomb scattering with remote continuum states in quantum dot devices} 
 
\author{R. Wetzler\email{wetzler@physik.tu-berlin.de}, A. Wacker, and 
E. Sch{\"o}ll} 
 
\affiliation{Institut f{\"u}r Theoretische Physik, Technische Universit{\"a}t
Berlin, Hardenbergstr.~36, 10623~Berlin, Germany} \date{\today}
 
\begin{abstract} 
Electron capture and emission by Coulomb scattering in self-assembled
quantum dot (QD) devices is studied theoretically. 
While the dependence of the Coulomb scattering
(Auger) rates on the local wetting layer electron density  has been
a topic of intense research, we
put special interest on the remote 
scattering between QD electrons and continuum
electrons originating from a quantum well, doped bulk layers or metal
contacts. Numerical effort is made to include all microscopic transitions
between the Fermi distributed continuum states. The remote Coulomb scattering
is investigated as a function of the electron  density, the distance from the QDs and
the temperature.  Our results are compared with experimental observations,
considering lifetime limitations in QD memory structures as well as the
electron emission in pn-diodes.
\end{abstract} 

\maketitle
 
\section{Introduction} 
The controversial discussions about the nature of the electron dynamics in
quantum dots \cite{bim99} (QDs) show that this topic is still not fully understood.
Different electron-phonon coupling regimes leading to a multi-phonon process
\cite{hei97,mag02a,fel01} or a polaron decay \cite{mag02,ver00} have been
proposed. Especially for a multi-phonon process a so-called bottleneck
effect \cite{boc90} is expected due to the discrete QD density of states.

On the other hand it has been shown theoretically \cite{boc92,usk98,usk97a} as
well as experimentally \cite{ray00,mor99b} that electron capture and
relaxation by Coulomb scattering (Auger effect) with electrons in the
wetting layer can be very efficient for large enough densities.
Furthermore, an Auger-type process can overcome the bottleneck due to the
continuous energy dispersion in the wetting layer.

The depletion of the wetting layer density by an electric field \cite{fry00a}
has been shown to reduce the relaxation rates. Similarly, the electron
density in the vicinity of the QDs is reduced in a pn diode by the depletion
layer within a DLTS experiment \cite{kap00a,wet00,wet03}. Electrons which are candidates
for a Coulomb scattering process are not located at the QDs anymore, but at
some distance.  Therefore we are interested in Coulomb scattering with
remote continuum electrons of a quantum well and bulk contact regions as a
function of the distance from the QDs.
 
We propose a theory to calculate the transition probabilities between two QD
and two continuum states. Debye screening effects are included according to
the three-dimensional Poisson equation. From the two-particle scattering
probability we calculate the total scattering probability of the electron
relaxation within a QD (Auger process) and the excitation from a lower state
into an excited state (impact ionization or impact excitation)  by a numerical integration over
the Fermi distributed continuum states.

\section{Theory} 
\subsection{General} 
\begin{figure} 
\includegraphics[angle=0,width=1.0\columnwidth]{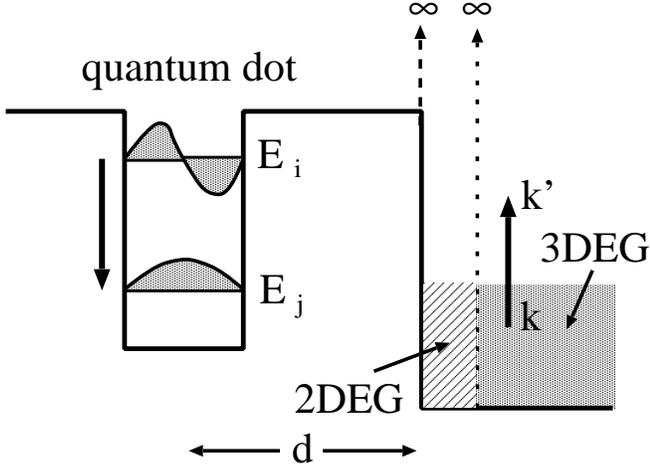}
\caption[a]{Sketched band diagram showing the remote Coulomb scattering with two- and three-dimensional continuum electrons} 
\label{fig1} 
\end{figure}

The band diagram of a QD and a remote quantum well and a bulk
region defined by the doping profile is sketched in Fig. \ref{fig1}. We assume
that the 2DEG and the 3DEG are confined by infinitely  high barriers.  The
transition probability per unit time between the QD states $i$ and $j$, and
the continuum states $\boldsymbol{k}$ and $\boldsymbol{k}'$ is calculated by
Fermi's Golden Rule
\begin{equation} 
W_{i\boldsymbol{k}\rightarrow j\boldsymbol{k}'}=\frac{2\pi}{\hbar}
|M_{i\boldsymbol{k}\rightarrow j\boldsymbol{k}'}|^2
\delta(E_{\text{ini}}-E_{\text{fin}})\text{,}
\end{equation} 
where $E_{\text{ini}}$ and $E_{\text{fin}}$ are the initial and final
energies. The matrix element is given by
\begin{equation} 
M_{i\boldsymbol{k}\rightarrow j\boldsymbol{k}'}=\iint\d^3\boldsymbol{r}
 \d^3\boldsymbol{r}'\psi^*_{\boldsymbol{k}'}(\boldsymbol{r})
 \psi^*_{j}(\boldsymbol{r}')U(\boldsymbol{r},\boldsymbol{r}')
 \psi_{i}(\boldsymbol{r}')\psi_{\boldsymbol{k}}(\boldsymbol{r}) \text{,}
\label{EqMatrixelement} 
\end{equation} 
where $\psi_{i}$, $\psi_{j}$, $\psi_{\boldsymbol{k}}$ and
$\psi_{\boldsymbol{k}'}$ symbolize the respective wave functions
\cite{lan91}. $U$ represents the effective Coulomb potential, which can be
written in a two-dimensional Fourier representation as
\begin{equation}
U(\boldsymbol{r},\boldsymbol{r}')=\frac{1}{A}\sum\limits_{\boldsymbol{q}'}U(q',z,z')e^{i\boldsymbol{q}'(\boldsymbol{r}_{\shortparallel}-\boldsymbol{r}'_{\shortparallel})}\text{,}
\end{equation}
with
\begin{equation}
U(q',z,z')=\frac{e^2e^{-q'|z-z'|}}{2q'\epsilon_0\epsilon_r\epsilon_e(q',z,z')}\text{.}
\end{equation}
Here, $\epsilon_r$ and $\epsilon_0$ are the relative and absolute
permittivities, $e>0$ is the elementary charge, $A$ is the normalization area,
$z$, $z'$ are  the coordinates in growth direction, $\boldsymbol{r}_{\shortparallel}$, $\boldsymbol{r}'_{\shortparallel}$ are the in-plane coordinates, and $\boldsymbol{q}'$ are two-dimensional wave vectors with
$|\boldsymbol{q}'|=q'$. The dielectric function $\epsilon_e$ accounts for
static screening effects of free electron densities. The evaluation of the
matrix elements (\ref{EqMatrixelement}) depends on the QD  wave
functions, which  themselves 
depend crucially on  the QD shape. In order to estimate the
integral, we note that the wave functions $\psi_{i}(\boldsymbol{r})$ and
$\psi_{j}(\boldsymbol{r})$ are orthogonal and thus the lowest order
contribution results from the dipole moment $-e{\bf a}$ of the effective
charge density $-e\psi^*_{i}(\boldsymbol{r})\psi_{j}(\boldsymbol{r})$. We
approximate this dipole moment by two localized charges of opposite sign and
thus obtain
\begin{equation} 
\label{eqn4} 
 \int\d^3\boldsymbol{r}'
\psi^*_{j}(\boldsymbol{r}')U(\boldsymbol{r},\boldsymbol{r}')
\psi_{i}(\boldsymbol{r}') \approx U(\boldsymbol{r},\frac{\boldsymbol{a}}{2})-
U(\boldsymbol{r},-\frac{\boldsymbol{a}}{2})\, \text{,}
\end{equation} 
where the center position of the QD is chosen as \mbox{$\boldsymbol{r}'=0$}.
The total transition probability per unit time from a QD state $i$ into a QD state $j$ is then obtained from an integration over the
continuum states
\begin{equation} 
W_{i\rightarrow
j}=\left(\frac{2\Omega}{(2\pi)^n}\right)^2\iint\d^n\boldsymbol{k}\d^n\boldsymbol{k}'
W_{i\boldsymbol{k}\rightarrow j\boldsymbol{k}'} f_k[1-f_{k'}]\text{,}
\end{equation} 
where $f_{k}$ and $f_{k'}$ are the Fermi distribution functions of the
initial and final continuum states. For a two-dimensional electron gas (2DEG) we
substitute $n=2$ and $\Omega=A$, and  $n=3$ and $\Omega=AL$ for a
three-dimensional electron gas (3DEG), where $L$ is a (large) normalization length.
\subsection{Two-dimensional electron gas} 
In order to find an expression for the dielectric function $\epsilon_e$ for a
two-dimensional electron gas positioned at $z=d$, we consider an elementary
charge positioned at $z=0$, corresponding to an external potential with the
Fourier components
\begin{equation}
U_{\text{ext}}(q,z)=\frac{e^2e^{-q|z|}}{2q\epsilon_0\epsilon_r}\text{.}
\end{equation}
For the induced potential $U_{\text{ind}}$ originating from the induced
electron density $\delta n_{\text{2D}}(q)$ in the quantum well layer the
Fourier components of the three-dimensional Poisson equation \cite{wet03} read as
\begin{equation}
-q^2U_{\text{ind}}(q,z)+\partial_z^2U_{\text{ind}}(q,z)=-\frac{e^2\delta
 n_{\text{2D}}(q)}{\epsilon_0\epsilon_r}\delta(z-d)\text{,}
\end{equation}
with the solution
\begin{equation}
\delta U_{\text{ind}}(q,z)=\frac{e^2}{2\epsilon_0\epsilon_rq}\delta
n_{\text{2D}}(q)e^{-q|z-d|}\text{.}
\end{equation}
Within the Debye screening approach, $\delta n_{\text{2D}}(q)$ is given by
\begin{equation}
\delta n_{\text{2D}}(q)=-\frac{\partial n_{\text{2D}}}{\partial E_{F}}
\left[U_{\text{ind}}(q,d)+U_{\text{ext}}(q,d)\right]\text{.}
\end{equation}
Assuming a single 2DEG quantization energy $E_{\text{2D}}$, $n_{\text{2D}}$ is
given by
\begin{equation}
n_{\text{2D}}=\frac{m^*k_BT}{\pi
\hbar^2}\ln\left(1+\exp\left(\frac{E_F-E_{\text{2D}}}{k_BT}\right)\right)\text{,}
\end{equation}
where $E_F$ is the Fermi energy, $m^*$ is the effective mass, $T$ is the temperature and $k_B$ is
Boltzmann's constant, leading to the dielectric function for a 2DEG
\begin{equation}
\label{defeps}
\epsilon_e^{-1}(q,d)=\frac{U_{\text{ext}}(q,d)+U_{\text{ind}}(q,d)}{U_{\text{ext}}(q,d)}=\left[1+\frac{\lambda_{\text{2D}}}{q}\right]^{-1}\text{,}
\end{equation}
with the inverse screening length
\begin{equation}
\lambda_{\text{2D}}=\frac{e^2m^*}{2\pi\epsilon_0\epsilon_r\hbar^2}\left[1+\exp\left(\frac{E_{\text{2D}}-E_F}{k_BT}\right)\right]^{-1}\text{.}
\end{equation}
Inserting the two-dimensional Fourier expansion of the Coulomb potential into
Eq.~(\ref{eqn4})  and using plane waves for the 2DEG states, we find
\begin{equation} 
\begin{split} 
&\left|M^{\text{2DEG}}_{i\boldsymbol{k}\rightarrow j\boldsymbol{k}'}\right|^2=
\left|\int\limits_A\frac{e^{-i\boldsymbol{k}'\boldsymbol{r}_{\shortparallel}}}{\sqrt{A}}\sum\limits_{\boldsymbol{q}'}\frac{e^2
e^{-q'd+i\boldsymbol{q}'\boldsymbol{r}_{\shortparallel}}}{2\epsilon_0\epsilon_r
A (q'_{\shortparallel}+\lambda_{\text{2D}})}\right.\\
&\left. \;\times\left[e^{q'\frac{a_z}{2}+i\boldsymbol{q}'\frac{\boldsymbol{a}_{\shortparallel}}{2}}
-e^{-q'\frac{a_z}{2}-i\boldsymbol{q}'\frac{\boldsymbol{a}_{\shortparallel}}{2}}\right]\frac{e^{i\boldsymbol{k}\boldsymbol{r}_{\shortparallel}}}{\sqrt{A}}
\d^2r_{\shortparallel}\right|^2 \text{.}
\end{split} 
\end{equation} 
$\boldsymbol{a}_{z}$ and $\boldsymbol{a}_{\shortparallel}$ are the $z$- and
parallel components of $\boldsymbol{a}$, $\boldsymbol{k}$ and
$\boldsymbol{k}'$ are the two-dimensional wave vectors of the initial and
final 2DEG states.
Integration in real space gives
\begin{equation} 
\label{M2D} 
\left|M^{\text{2DEG}}_{i\boldsymbol{k}_{\shortparallel}\rightarrow
j\boldsymbol{k}'_{\shortparallel}}\right|^ 2=\frac{ e^4e^{-2q d}\left[\cosh(q
a_z)-\cos(\boldsymbol{q}\boldsymbol{a}_{\shortparallel})\right]}{2 A^2
(\epsilon_0\epsilon_r)^2(q+\lambda_{\text{2D}})^2}\text{,}
\end{equation} 
with the momentum transfer $q=|\boldsymbol{q}|
=|\boldsymbol{k}-\boldsymbol{k}'|$.
\subsection{Three-dimensional electron gas} 
Coulomb scattering with semi-confined bulk electrons located at $z>d$ is
considered next. The Fourier components of the three-dimensional Poisson
equation for this problem are
\begin{equation}
\label{Poisson3D}
-q^2U_{\text{ind}}(q,z)+\partial_z^2U_{\text{ind}}(q,z)=-\frac{e^2\delta
 n_{\text{3D}}(q,z)}{\epsilon_0\epsilon_r}\Theta(z-d)\text{,}
\end{equation}
where the induced bulk electron density is given by
\begin{equation}
\delta n_{\text{3D}}(q,z)=-\frac{\partial n_{\text{3D}}}{\partial E_{F}}
\left[U_{\text{ind}}(q,z)+U_{\text{ext}}(q,z)\right]\text{.}
\end{equation}
With the definition of the Debye screening wave vector
\begin{equation}
\lambda_{\text{3D}}=\sqrt{\frac{e^2}{\epsilon_0\epsilon_r}\frac{\partial
n_{\text{3D}}}{\partial E_F}}
\end{equation}
and the expression for the bulk electron density
\begin{equation}
n_{\text{3D}}=2\left(\frac{m^*k_BT}{2\pi\hbar^2}\right)^{3/2}F_{1/2}\left(\frac{E_F-E_C}{k_BT}\right)
\end{equation}
the Debye screening wave vector is given by
\begin{equation}
\lambda_{\text{3D}}=\sqrt{\frac{e^2}{\epsilon_0\epsilon_r}n_{\text{3D}}
\frac{F_{-1/2}\left(\frac{E_F-E_C}{k_BT}\right)}{F_{1/2}\left(\frac{E_F-E_C}{k_BT}\right)}}\text{.}
\end{equation}
Here, $F_s$ denotes the Fermi integral of order $s$. Solving
Eqn.~(\ref{Poisson3D}) leads to
\begin{equation}
U_{\text{ind}}(q,z>d)=\frac{-e^2e^{-qz}}{2q\epsilon_0\epsilon_r}+\frac{e^2e^{-qd+\sqrt{\lambda^2_{\text{3D}}+q^2}(d-z)}}{\epsilon_0\epsilon_r\left[q+\sqrt{\lambda^2_{\text{3D}}+q^2}\right]}\text{.}
\end{equation}
Here, we have used that $U_{\text{ind}}$ is continuously differentiable at
$z=d$. Analogously to Eqn. (\ref{defeps}) we find the dielectric function for
the inhomogeneous bulk electron distribution for $z>d$ as
\begin{equation}
\begin{split}
\epsilon_e^{-1}(q,z>d)=\frac{2qe^{-(z-d)(\sqrt{\lambda_{\text{3D}}^2+q^2}-q)}}{q+\sqrt{\lambda_{\text{3D}}^2+q^2}}\text{.}
\end{split}
\end{equation}
Using continuum wave functions, which are semi-confined in the $z$-direction
and plane waves in the radial directions then leads to
\begin{equation} 
\begin{split} 
&\left|M^{\text{3DEG}}_{i\boldsymbol{k}\rightarrow
j\boldsymbol{k}'}\right|^2= \frac{1}{A^2}\left|\int\limits_d^{\infty}\d
z\sqrt{\frac{2}{L}}\sin(k'_z(z-d))\right.\\
&\times\frac{e^2}{\epsilon_0\epsilon_r}\frac{e^{-qd+\sqrt{\lambda_{\text{3D}}^2+q^2}(d-z)}}{q+\sqrt{\lambda_{\text{3D}}^2+q^2}}\left[e^{q\frac{a_z}{2}+i\boldsymbol{q}\frac{\boldsymbol{a_{\shortparallel}}}{2}}-e^{-q\frac{a_z}{2}-i\boldsymbol{q}\frac{\boldsymbol{a_{\shortparallel}}}{2}}\right]\\
&\;\;\;\;\;\;\;\;\;\;\;\;\;\;\;\;\;\;\;\;\;\;\;\;\;\;\times
\left.\sqrt{\frac{2}{L}}\sin(k_z(z-d))\right|^2\text{,}
\end{split} 
\label{EqMquadrat3d}
\end{equation} 
where $k_{z}$ and $k'_{z}$ denote the $z$-components of the wave vectors.
Integration over real space results in
\begin{equation}
\label{M3D} 
\begin{split} 
&\left|M^{\text{3DEG}}_{i\boldsymbol{k}\rightarrow
j\boldsymbol{k}'}\right|^2= \frac{2 e^4e^{-2q d}\left[\cosh(q
a_z)-\cos(\boldsymbol{q}\boldsymbol{a}_{\shortparallel})\right]}{(AL)^2
(\epsilon_0\epsilon_r)^2}\\
&\frac{\lambda_{\text{3D}}^2+q^2}{(\sqrt{\lambda_{\text{3D}}^2+q^2}+q)^2}
\left[\frac{1}{p_z^2+\lambda_{\text{3D}}^2+q^2}-\frac{1}{q_z^2+\lambda_{\text{3D}}^2+q^2}\right]^2
\end{split} 
\end{equation} 
with $q_{z}=k'_z-k_z$ and $p_{z}=k'_z+k_z$. Analogously to the 2DEG, the
transition probability is obtained by integration over occupied continuum
states, which is performed numerically.

\section{Results} 
\subsection{Remote two-dimensional electron gas}
\begin{figure} 
\includegraphics[angle=0,width=1.0\columnwidth]{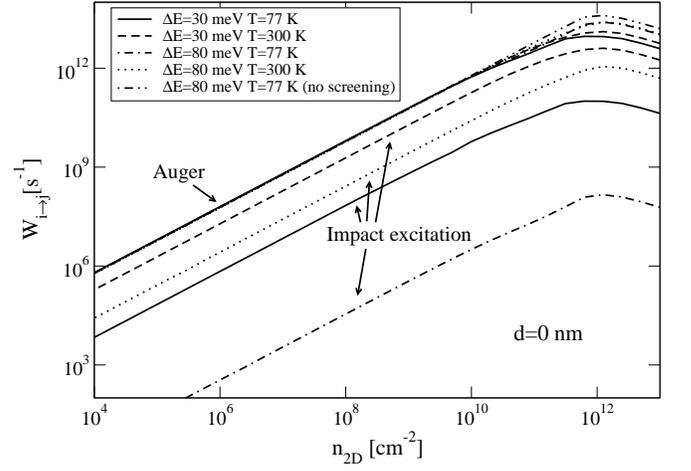}
\caption[a]{Auger and impact  excitation probabilities as a 
function of the 2DEG electron density at $d=$0 nm  (e.g., for 
a wetting layer surrounding the QDs)
for different energy differences $\Delta E$ and temperatures $T$.
In addition the corresponding result without screening is shown
for the Auger rate at  $\Delta E=$80 meV and 77 K (dash-double-dotted line).
}
\label{fig2} 
\end{figure} 
According to the geometry and size of InAs QDs which have a typical base
length of about 10 nm and  a strong quantization in z-direction, we use
$|\boldsymbol{a}_{\shortparallel}|$=5 nm and $a_z$=0. For small dipole
moments a Taylor expansion in Eqn. (\ref{M2D}) and (\ref{M3D}) leads to a
quadratic dependence on the transition dipole moments, and the transition
probabilities presented here can be rescaled for specifically shaped QDs.

First we performed calculations of the Auger and impact excitation
probabilities per unit time as a function of the electron density per unit area in the
wetting layer, using $d=$ 0 nm and $m^*_{\text{InAs}}=$ 0.023 $m_0$ for
different QD level spacings $\Delta E=|E_j-E_i|$. Fig. \ref{fig2} shows the
resulting probabilities for $\Delta E=$30 meV and 80 meV for T=77 K and 300 K.
The Auger probabilities hardly depend on $\Delta E$ and the temperature. For
densities less than $10^{10}$ cm$^{-2}$ the Auger probability is a linear
function of the density. From a fit $W_{i\rightarrow j}=T_{\text{Aug}}^{\text{2D}}n_{\text{2D}}$ for low densities we obtain an Auger
relaxation coefficient of $T_{\text{Aug}}^{\text{2D}}=$60 cm$^2$s$^{-1}$.
This value is about one  order of magnitude larger than the corresponding
value calculated in Ref. \cite{usk98} (1 cm$^2$s$^{-1}$). We explain this
deviation by the different values for the transition dipole moment and the
effective mass used in this work. 
 Note that the Auger capture
coefficients  evaluated in Ref. \cite{usk98} (2$\times$10$^{-12}$
cm$^4$s$^{-1}$) are also much smaller than those observed
experimentally in  Ref. \cite{ray00} (10$^{-8}$ cm$^4$s$^{-1}$).
For large densities, i.e., when the 2DEG becomes degenerate, the density
dependence becomes nonlinear, and the probabilities are even reduced.
 
Fig. \ref{fig2} also shows that the impact excitation probabilities strongly
depend on $\Delta E$ and $T$, and the ratio between the impact excitation and
Auger probabilities is given by the Boltzmann factor $\exp(-\Delta E/(k_BT))$.
The Auger probability for  $\Delta E=$80 meV and T=77 K but neglecting
screening effects is also plotted in Fig. \ref{fig2}, demonstrating that
screening effects play only a role for densities typically larger than
$10^{10}$ cm$^{-2}$.
 \begin{figure} 
\includegraphics[angle=0,width=1.0\columnwidth]{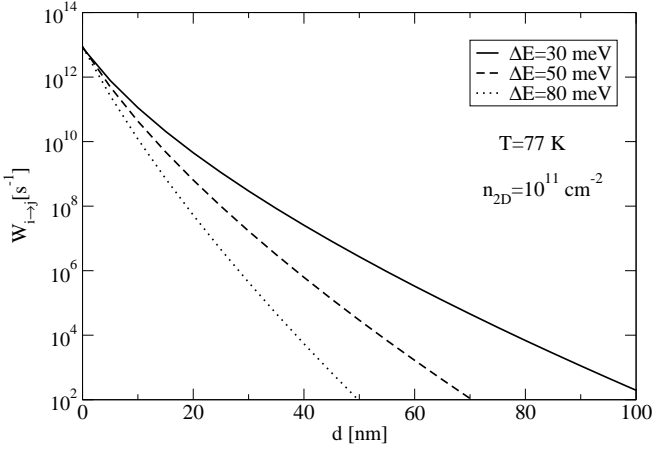}
\caption[a]{Remote Auger probabilities as a function of the distance of the 2DEG from the QDs for $n_{\text{2D}}=$ 10$^{11}$ cm$^{-2}$, $\Delta E=$30 meV, $\Delta E=$50 meV and
 $\Delta E=$80 meV at 77 K }
\label{fig3} 
\end{figure} 

Fig. \ref{fig3} shows the remote Auger probabilities as a function of the
distance $d$ between the QDs and the 2DEG, for $n_{\text{2D}}=$ 10$^{11}$
cm$^{-2}$, $m^*_{\text{GaAs}}=$ 0.06 $m_0$, $T=$ 77 K and $\Delta E=$30 meV,
$\Delta E=$50 meV and $\Delta E=$80 meV. The Auger probabilities decrease
exponentially with $d$, depending on the energy transfer. Therefore the
Coulomb scattering of this type has a local character. As a consequence, a
2DEG at $d=$200 nm as in the memory structure of Ref. \cite{yus97} does not
contribute to the lifetime limitation of charged QDs.

\subsection{Remote three-dimensional electron gas} 
 \begin{figure} 
\includegraphics[angle=0,width=1.0\columnwidth]{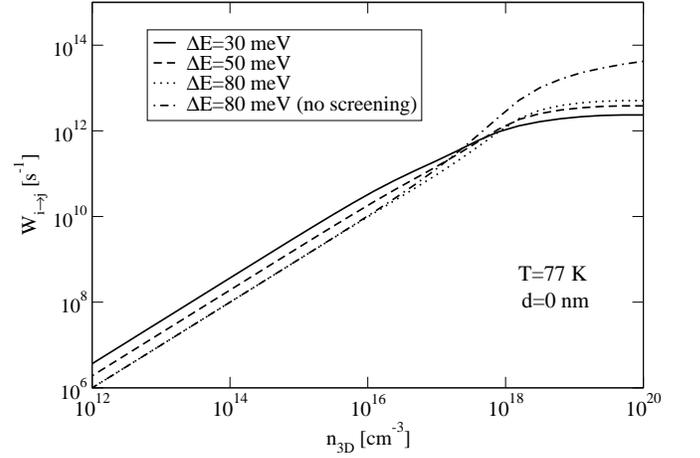}
\caption[a]{Auger probabilities as a function of the three-dimensional electron density  for
 different energy level separations $\Delta E$
at 77K and with $d=$0 nm.
In addition the corresponding result without screening is shown
for the Auger rate at  $\Delta E=$80 meV  (dash-dotted line).
}
\label{fig4} 
\end{figure} 
Next, we calculated the Coulomb scattering probabilities as a function of the
bulk density in the vicinity of the QDs. The result is plotted for $\Delta
E=$30 meV (solid line), $\Delta E=$50 meV (dashed line) and $\Delta E=$80
meV (dotted line) in Fig. \ref{fig4} for T=77 K, $d=$0 nm and
$m^*_{\text{GaAs}}=$ 0.06 $m_0$. Again, for low densities the probabilities
are linear in the density. But in comparison to the 2DEG, the fitted Auger
coefficients $T_{\text{Aug}}^{\text{3D}}=W_{i\rightarrow j}/n_{\text{3D}}$ depend on the QD level spacings. Here we obtain
$T_{\text{Aug}}^{\text{3D}}$=$4.7\times$10$^{-6}$ cm$^3$s$^{-1}$,
$T_{\text{Aug}}^{\text{3D}}$=$1.9\times$10$^{-6}$ cm$^3$s$^{-1}$ and
$T_{\text{Aug}}^{\text{3D}}$=$1.0\times$10$^{-6}$ cm$^3$s$^{-1}$ for $\Delta
E=$30 meV, $\Delta E=$50 meV and $\Delta E=$80 meV, respectively. For high
electron densities we observe a deviation from this linear behavior, since
pairs of occupied and empty continuum states can only be found near the Fermi
level when the electron gas becomes highly degenerate. The resulting Auger
probability neglecting screening effects for $\Delta E=$80 meV is plotted as a
dash-dotted line in Fig. \ref{fig4}, demonstrating that screening is only
important for densities above 10$^{17}$ cm$^{-3}$.

 \begin{figure} 
\includegraphics[angle=0,width=1.0\columnwidth]{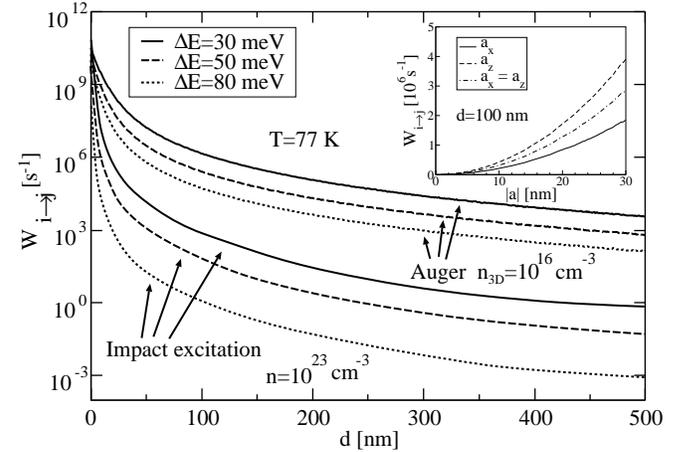}
\caption[a]{The Auger probabilities for a three-dimensional electron gas with  $n_{\text{3D}}=$10$^{16}$ cm$^{-3}$, 
$m^*=$0.06 $m_0$ and the impact  excitation probabilities for 
$n_{\text{3D}}=$10$^{23}$ cm$^{-3}$, $m^*=m_0$ as a function of $d$ 
for  different level spacings $\Delta E$ at 77 K.
Inset: dependence of the Auger probability 
for  $n_{\text{3D}}=$10$^{16}$ cm$^{-3}$, 
$m^*=$0.06 $m_0$ and $\Delta E=80$ meV on the transition dipole 
moment for
$d$=100 nm: $a_x\neq 0$, $a_z=0$ (solid line); $a_x=0$, $a_z\neq 0$
(dashed line) and $a_x$=$a_z\neq 0$ (dash-dotted line)}
\label{fig5} 
\end{figure} 
The Auger probability as a function of the distance from the QDs is
investigated next. The resulting curves for $n_{\text{3D}}=$10$^{16}$
cm$^{-3}$, $m^*_{\text{GaAs}}=$0.06 $m_0$, $\Delta E=$30 meV (solid line),
$\Delta E=$50 meV dashed line, $\Delta E=$80 meV (dotted line) at 77 K are
shown in Fig. \ref{fig5}. Compared to the remote 2DEG scattering, the
dependence on $d$ is weaker, since momentum in the 3DEG can additionally
be changed in the third dimension. 
An interesting point is how a metal
contact with its high electron density limits the lifetime of charged QDs.
Therefore, the impact excitation probabilities are also displayed in
Fig. \ref{fig5} for $n_{\text{3D}}=$10$^{23}$ cm$^{-3}$, which is a typical
electron density in metal contacts, and $m^*=$$m_0$. Depending on the energy
spacing $\Delta E$, we obtain transition times of the order of an hour for
$\Delta E=$80 meV and $d=$400 nm. This goes well with the observation
of bistable behavior in QD structures on this time scale \cite{yus97,rac02}.
The inset of Fig. \ref{fig5} shows the dependence of
the Auger probability on the transition dipole moment $-e \boldsymbol{a}$,
 which can be well approximated by a quadratic law 
resulting from the Taylor expansion of Eq.~(\ref{EqMquadrat3d}).
The Taylor expansion  also explains that the  transition probabilities
are enlarged by a factor of 2 for 
vertical dipole moments (in growth direction)
compared to the lateral dipole moments 
studied throughout this work.
We conclude that the direction of
the dipole moment does not affect the order of magnitude of the
transition rates.

 \begin{figure} 
\includegraphics[angle=0,width=1.0\columnwidth]{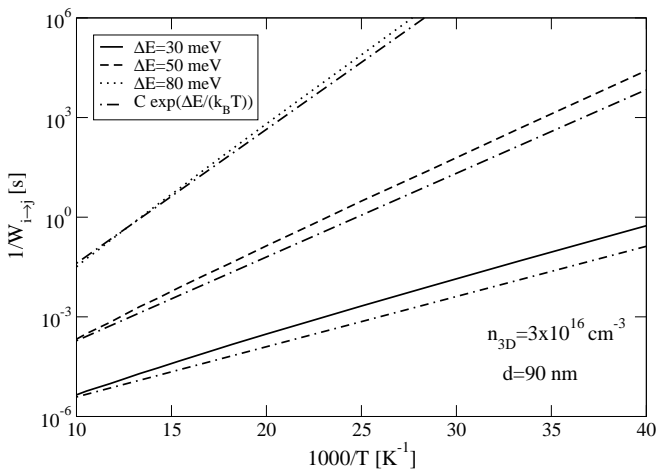}
\caption[a]{Arrhenius plot of the inverse impact excitation probability with $d$=90 nm, $n_{\text{3D}}=$3$\times$10$^{16}$ cm$^{-3}$ for $\Delta E=$30 meV (solid line), $\Delta E=$50 meV dashed line, $\Delta E=$80 meV (dotted line); dash-dotted lines: functions proportional to the Boltzmann factor  $C \exp(\Delta E/(k_BT))$ where $C$ is chosen such that
lines intersect at 1000/$T$=10 K$^{-1}$} 
\label{fig6} 
\end{figure}

Finally we consider QDs embedded in a pn diode, where the QDs are positioned
within 3$\times$10$^{16}$ n-doped GaAs layers. Such a device was investigated 
in  DLTS experiments, and a level spacing of $\Delta E=$82 meV with an
emission time of 62 ms at T=40 K has been observed \cite{kap00a}. The
emission from the QD ground state level is dominated by an excitation into the
excited state, from which the electron escapes into the continuum by a fast
tunneling process.  Taking into account  the Boltzmann factor
between excitation and relaxation  processes 
we estimate a timescale for the relaxation process from the excited state into the ground state
of about 3 ps. 
From Fig. \ref{fig2} we conclude that for an Auger process on
that fast time scale a wetting layer density of 10$^{10}$ cm$^{-2}$ 
 would be
necessary. However, the wetting layer is not occupied due to the electric
field in the depletion layer. From the parabolic band bending and a QD ground
state energy of 200 meV below the conduction band edge we estimate that the
depletion layer extends to $d=$90 nm from the QDs. For these parameters, we
display the temperature dependence as an Arrhenius plot as shown in
Fig. \ref{fig6} for $\Delta E=$30 meV (solid line), $\Delta E=$50 meV dashed
line, $\Delta E=$80 meV (dotted line). The Botzmann factor (functions
proportional to $\exp(\Delta E/(k_BT))$ chosen such that the lines intersect at
1000/$T$=10 K$^{-1}$) is also displayed as dash-dotted line for each energy
spacing, showing that the slope of the Arrhenius plot of the
temperature dependence is mainly given by the energy spacing.

Nevertheless, we obtain an impact excitation time of about 10$^4$ s for $T=$40
K and $\Delta E=$80 meV, which is orders of magnitudes slower than observed in
the DLTS experiment. Only by reducing the level spacing 
below $\Delta E=$50 meV  impact excitation  times in the range of tens of ms
can be obtained at this temperature. 
This indicates that in addition phonon mechanisms have to be taken into account.

\section{Conclusion} 
In conclusion, we have  evaluated the Auger effect and impact
 excitation (Coulomb scattering) in QD devices with 
remote continuum states.
Three-dimensional Debye screening has been included, which is important for
high electron densities.  Our results show that, depending on the detailed device and QD geometries, scattering with remote 
continuum carriers can be of importance for the electron kinetics
 if the local wetting layer density is reduced.

This work was supported by DFG in the framework of Sfb 296. Helpful
discussions with M. Geller and R. Heitz are acknowledged.

\end{document}